\documentclass{jaa}
\usepackage{natbib}
\usepackage{graphicx}
\newcommand{\mnras}   {{ MNRAS}}
\newcommand{\apj}   {{ Astrophysical Journal}}
\newcommand{\aap}   {{Astronomy and Astrophysics}}
\newcommand{\aj}   {{Astronomical Journal}}
\newcommand{\araa}   {{Annual Review of Astron and Astrophys}}

\usepackage{hyperref, geometry}

\begin{document}\sloppy

\title{$U\!BV\!I$  CCD~Photometry of Berkeley 55 Open Cluster}

\author{{\.I}nci Akkaya Oralhan\textsuperscript{1,*}}

\affilOne{\textsuperscript{1}Department of Astronomy and Space Sciences, Science Faculty, Erciyes University,TR-38039, Kayseri, Turkey.}

\twocolumn[{

\maketitle
\corres{iakkaya@erciyes.edu.tr}
\msinfo{26 January 2021}{26 January 2021}

\begin{abstract}
Fundamental astrophysical parameters have been derived for Be~55 open cluster based on  $U\!BV\!I$~CCD photometric data, observed with the AZT-22 1.5m telescope at Maidanak Astronomical Observatory in Uzbekistan. The mean reddening is obtained as E(B-V)=1.77$\pm$0.10 mag from early type members. The zero age main sequence fitting in the $Q_{V\lambda}$--$Q'$ diagrams indicates the distance modulus, $(V_{0}$--$M_{\rm V}$)=12.4$\pm$0.20~mag~(d$=$3.02$\pm$0.28~kpc). This photometric distance is consistent with the distances of Gaia EDR3 (d$=$3.09$\pm$0.16~kpc) and period-luminosity relation (d$=$2.78$\pm$0.32~kpc) of its Cepheid $S5$ within the uncertainties. This distance also locates the cluster near the Perseus spiral arm. The Geneva isochrone fittings to the Hertzsprung-Russell diagram and observational colour-magnitude diagrams derive turn-off age, 85$\pm$13~Myr, by taking care five red supergiants/bright giants. The possible inconsistences on the locations of the bright giants with the rotating/non-rotating isochrones may be due to both the age spread of stars in young open clusters and the diversity in rotational velocities.
\end{abstract}

}]

\doinum{12.3456/s78910-011-012-3}
\artcitid{\#\#\#\#}
\volnum{000}
\year{0000}
\pgrange{1--}
\setcounter{page}{1}
\lp{1}

\section{Introduction}

Be~55 was studied by \cite{neg12} (hereafter N12), \cite{loh18} (hereafter L18), and \cite{alo20} (hereafter A20). N12 identified populations of B-type stars, one F-type and four late-type red supergiants/bright giants (RSGs/RBGs), two Blue Stragglers (BS). Their spectroscopic and photometric analyses indicated an age of 50$\pm$10 Myr and a distance of d$\approx$4 kpc. 
Its member $S5$ was identified by L18 as a Type I Cepheid with a pulsation period 5.85 day. Spectral types, effective temperatures and radial velocities of these bright giants have been obtained by A20 from the medium-resolution spectroscopic observations of the 4.2 m William Herschel Telescope at La Palma.  They updated the distance and age of Be~55 as $3.24\pm0.22$ kpc and $63\pm15$ Myr, respectively from PARSEC isochrones \citep{bre12}.
Be~55 were also studied by \cite{mac07}, \cite{tad08}, \cite{buk11}, and \cite{mol18}, respectively (See Table 7).

As emphasized by A20, Cepheids in young open clusters (OCs) which lie on the blue loop as pulsating variable, have great importance for distance estimation. For their distances, there are useful  period-luminosity relations (PLR) in the literature \citep{ben07,and13,laz20}. 
Some young OCs with Cepheids as an example are as the following, van den Bergh-Hagen \citep{mar14}, Be~51 \citep{neg18}, NGC~6649 and NGC~6664 \citep{alo20}.

Young OCs are generally heavily obscured. For their reddenings, $Q$ technique from their early-type stars is used \citep{sun13} (hereafter S13). For their bright evolved stars, the intrinsic colours of \cite{fer63} and \cite{fit70} for (UBV) and \cite{koo83} for JHK$_{s}$ are utilised.

In this paper new Maidanak $U\!BV\!I$ CCD~Photometry of Be~55 is presented, and analysed by utilising Gaia EDR3 astrometric/photometric data \citep{bro20,lin20}. The issues mentioned above have been studied. As emphasized by \cite{lin20}, compared to Gaia DR2 \citep{bro18,lin18}, the average improvement on the uncertainties of parallax/proper motion data of Gaia EDR3 is almost a factor 0.8 for the parallaxes, and 0.5 for the proper motions.

This paper is organized as follows: Section~2 describes the observation and data reduction. The radius, membership selection, the determination of the reddening, and distance modulus/distance are presented in Section 3. Be~55's age plus kinematics/orbital parameters of its bright giants are given in Sections 4--5.  A Discussion and Conclusion is presented in Section~6.

\begin{figure}[!t]
	\centering{\includegraphics[width=6cm, height=6cm] {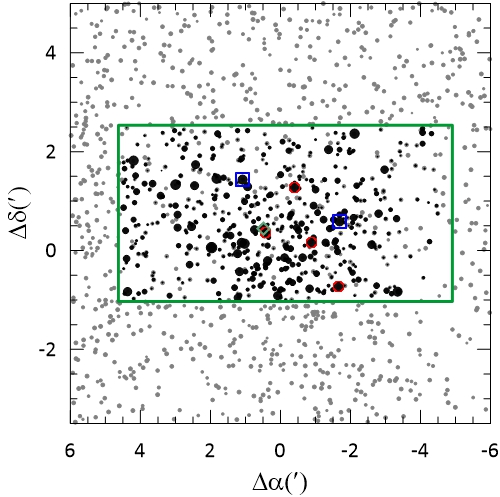}} \hspace*{1mm}
	\caption{The star chart of Be~55 based on Gaia EDR3 data (grey dots) and Maidanak 2k data (black dots). The position of the stars is relative to its mean equatorial coordinates. Green rectengular shows the field of view of Maidanak detector. Five RSGs/RBGs and two BSs are also indicated (see Fig.~6 for the symbols).}
	\end{figure}

\begin{table*}[!t]\label{Table-1}
\centering
\caption{For $UBVI$ filters, exposure time, $FWHM$-seeing, the primary/secondary extinction coefficients ($k_{1\lambda}$,  $k_{2\lambda}$), and the photometric zero-point ($\zeta_{\lambda}$).}
\begin{tabular}{ccccccc}
	\hline
	Filter  &Colour& Exposure time~(s) &$FWHM$-seeing~$(^{\prime\prime})$&$k_{1\lambda}$ &$k_{2\lambda}$&$\zeta_{\lambda}$ \\
	\hline	
	U &U-B &600      &1.22 &0.423$\pm$0.022  &0.023 &21.560$\pm$0.011 \\
	B &B-V &20,~600  &1.99 &0.324$\pm$0.010 &0.026 &23.261$\pm$0.006 \\
	V &B-V &10,~300  &0.88 &0.231$\pm$0.009  &  -   &23.396$\pm$0.001 \\
	I &V-I &10,~120  &0.74 &0.139$\pm$0.010  &  -   &23.017$\pm$0.009\\
		\hline
\end{tabular}
\end{table*}

\begin{table*}
	\centering
	\caption{Photometric data of Be~55.}
	\begin{tabular}{lccccccccc}
		\hline
		$\alpha_{2000}$ (h\,m\,s) &$\delta_{2000}$ $(^{\circ}\,^{\prime}\,^{\prime\prime})$  &V-~mag&  $(U$--$B)$ & $(B$--$V)$ & $(V$--$I)$ & eV & e$(V$--$I)$ & e$(B$--$V)$ & e$(U$--$B)$\\
		\hline
		21 16 51.27 &51 46 07.3&15.186&0.524&1.562&2.130&0.003&0.011&0.004&0.009 \\
		21	17	2.30  &	    51	46	58.1 &14.666&0.581&1.543&2.017&0.001&0.001&0.001&0.006 \\
		21	17	1.72  &  51	 46	49.1 &15.771&0.721&1.532&2.021&0.001&0.005&0.004&0.016\\
		... &...&...&...&...&...&...&...&...&...\\
		... &...&...&...&...&...&...&...&...&...\\
		... &...&...&...&...&...&...&...&...&...\\
		\hline
	\end{tabular} 
\end{table*}

\section{Observation and Data Reduction}

The observation in Johnson-Cousins' $U\!BV\!I$ system of Be 55 was carried out with the AZT-22 1.5-m (f/7.74) Ritchey-Chretien telescope at Maidanak Astronomical Observatory (MAO) in Uzbekistan, during high-quality photometric night, on 2005 Agust 9.  Images were taken with the {\it SITe~2k} CCD detector, which has 2000$\times$800 pixels, a gain of 1.16 e$^-$/ADU and a readout noise of 5.3 e$^-$. The combination of the telescope and the detector ensures an unveignetted field of view of $8.85^{\prime}\times3.55^{\prime}$.  Standard magnitude for a given filter $\lambda$ is obtained using the following relation,

\begin{equation}
M_{\lambda} = m_{\lambda} - [k_{1\lambda} -k_{2\lambda}C)] X + \eta_{\lambda} C + \zeta_{\lambda}
\end{equation} 

where $m_{\lambda}$, $k_{1\lambda}$, $k_{2\lambda}$, $C$, and $X$  are the observed instrumental magnitude, primary/secondary extinction coefficients, colour index and air mass, respectively. $M_{\lambda}$, $\eta_{\lambda}$, $\zeta_{\lambda}$ are standard magnitude, transformation coefficient and photometric zero point, respectively. The other details of data reduction can be found in \cite{lim09}. The SAAO standard stars in \cite{men91} and \cite{kil98} were used to derive atmospheric extinction and transformation coefficients. Pre-processing was performed using the IRAF/ CCDRED package and an aperture of 10$^{\prime\prime}$ was used for standard star photometry. Exposure time~(s), \textit{FWHM}- seeing~$(^{\prime\prime})$, extinction coefficients and zero points for $U\!BV\!I$ filters are given in Table 1. Here, $FWHM$-seeing means the seeing, which is estimated from full width at half maximum of the point like stars on the images.
The photometric data of 357 stars are listed in Table 2 for a sample data.

\begin{table}
	\begin{center}
		\caption {The mean photometric errors of $<V>$~mag, $<(V-I)>$, $<(B-V)>$ and $<(U-B)>$ against $V$-mag.}
		\label{tab:errors}
		\setlength{\tabcolsep}{0.20cm}
		{\scriptsize
			\begin{tabular}{cccccc}
				\hline
				
				$V$ &$<\sigma_{V}>$&$<\sigma_{V-I}>$&$<\sigma_{B-V}>$&$<\sigma_{U-B}>$\\
				\hline
				13-14 &0.003 &0.013 &0.008 &0.005 \\
				14-15 &0.005 &0.020 &0.007 &0.016 \\
				15-16 &0.008 &0.007 &0.007 &0.017 \\
				16-17 &0.005 &0.007 &0.009 &0.032 \\
				17-18 &0.006 &0.009 &0.010 &0.065 \\
				18-19 &0.012 &0.016 &0.023 &0.129\\
				19-20 &0.016 &0.021 &0.053 &-  \\
				20-21 &0.050 &0.052 &0.123 & - \\	
				21-22 &0.090 &0.095 &0.156 &-\\
				22-23 &0.150 &0.157 &	-  &-\\
				
				\hline
			\end{tabular}
		} 
	\end{center}
\end{table}

Be~55'star chart is displayed in Fig.~1.  Its mean central equatorial (J2000) and the Galactic coordinates are as the following, RA =21$^{h}$ 16$^{m}$ 58$^{s}$ .01, Dec = +51$^{\circ}$ 45$^{\prime}$32$^{\prime\prime}$.04; $\ell$=93$^{\circ}$.027, $b$=1$^{\circ}$.798, respectively.

The mean photometric errors of $V$~mag, the colour indices $(V-I)$, $(B-V)$ and $(U-B)$ against $V$-mag for Be~55 are presented in Table~3. The photometry for Be~55 is compared to the $UBV~CCD$ photometry of N12.  For 26 common stars, the differences of $\Delta V$ and $\Delta (B-V)$ against $(B-V)$, and the difference of $\Delta (U-B)$ against $(U-B)$ are displayed in Fig.~2. When the large deviant data of the evolved stars are excluded from the calculation, the mean differences together dispersions of $V$ and $(B-V)$ are $\Delta V=+0.036\pm0.027$~mag and $\Delta (B-V)=+0.003\pm0.012$, respectively. The difference, $\Delta (U-B)$ is +0.082$\pm$0.064, which is systematically bluer than N12. For the interval of  $14 < V < 18$~mag, the $V$ magnitudes of N12 seem to be slightly fainter. For $14 < V < 18$~mag, $(B-V)$ values of this paper are in good consistent with N12, except for bright giants. In panel (c) there appears a discrepancy between -0.4  to -1.0 in $(U-B)$ of four RBGs/RSGs between N12 and Maidanak observations. For Cepheid S5 (red diamond), the discrepancy is about 0.50~mag in $\Delta V$ and 0.2~mag in $\Delta (B-V)$ from panels~(a) and (b). The reason for large colour difference for four late-type supergiants with very red colours between $(U-B)$ of present photometry and N12 is due to the so-called red leak in the used $U$ filter. Therefore, their $U$-magnitudes of present photometry are brighter than N12. However, there is no any red leak effect in the used $U$ filter, as is seen from table~2 of the transformation coefficients of \cite{lim09}.

\begin{figure}[!t]\label{F-figures-03}
	\centering{\includegraphics[width=0.7\columnwidth]{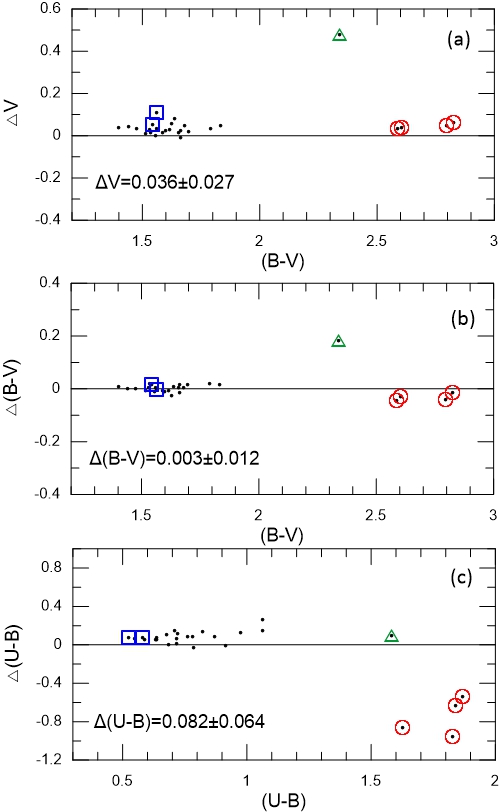}}
	\caption {The differences of $V$~mag, $(B-V)$ and $(U-B)$ as a function of $(B-V)$ and $(U-B)$. $\Delta$ means this paper - N12.  Five RSGs/RBGs are plotted for just presentation (for the symbols, see Fig.~6).}
\end{figure}

\section{Fundamental Parameters of Be 55}

\subsection{Radius of Be 55}

For the size of Be~55, its stellar radial density profile (RDP) (Fig.~3) is constructed from the Gaia EDR3 photometric/astrometric data for the field and cluster members within $15'.0$, down to $G =20$ mag. Its RDP has been constructed by counting stars in concentric rings of increasing width with distance to its centre. The number and width of rings were optimised so that the resulting RDP had adequate spatial resolution with moderate 1$\sigma$ Poisson errors \citep{bon07}. The solid curve (Fig.~3) denotes the fitted King's profile \citep{kin66}. The two-parameter function, $\sigma(R) =  \sigma_{bg} + \sigma_0/(1+(R/R_c)^2)$ is adopted. Here, $\sigma_{bg}$ is the residual background density, $\sigma_0$ the central density of stars, and R$_{core}$ the core radius. The horizontal gray bar shows the stellar background level measured in the comparison field. From Fig.~3, the core and cluster radii have been determined as $(R_{core},~R_{RDP})$ = $(0'.45,~2'.5)$. The cluster radius is compatible with $2'.5$ of \cite{mol18} and $2'.2$ of \cite{sam17}. $R_{core}$ value of this paper is less than  $1'.32$ of \cite{buk11}, $0'.90$ of \cite{kha13}, and $1'.28$ of A20, respectively.

\begin{figure}
	\centering{\includegraphics[width=8cm, height=6cm] {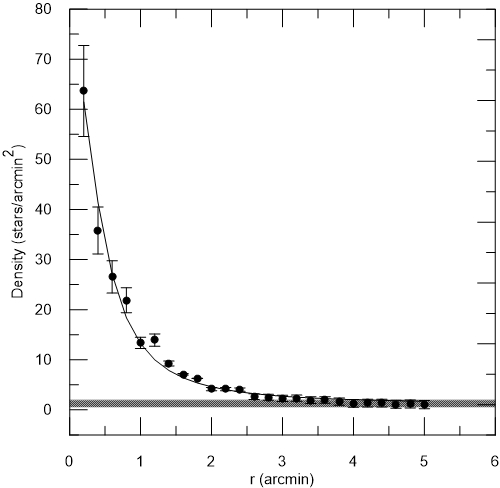}} 
	\caption{The radial density profile of Be~55. The curved line shows the fitting of King (1966). The vertical bars denote the Poisson errors.}
\end{figure}

\begin{figure}[!t]\label{Fig-4}
	\centering{\includegraphics[width=0.45\textwidth]{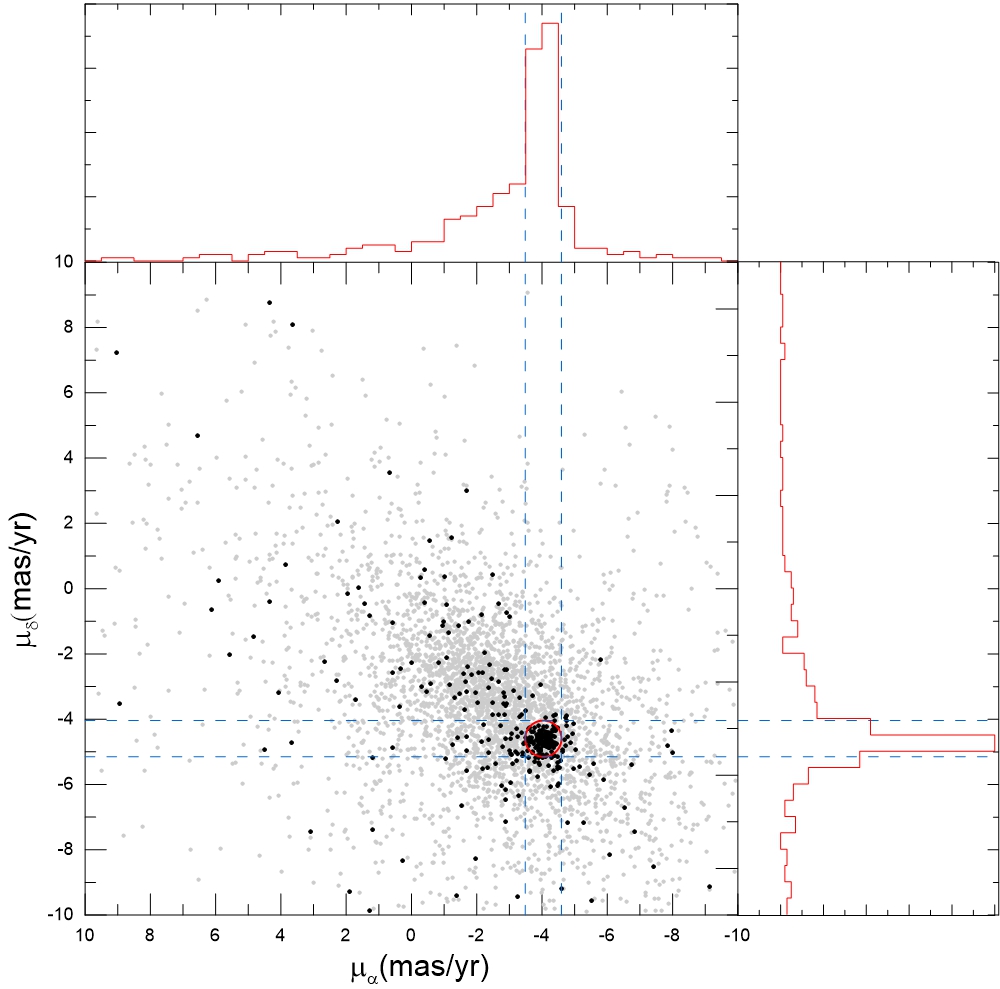}} \
	\centering{\includegraphics[width=0.46\columnwidth]{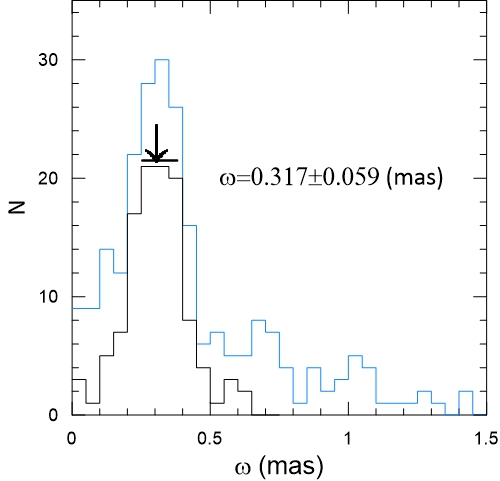}} \hspace*{1mm}
	\includegraphics[width=0.5\columnwidth]{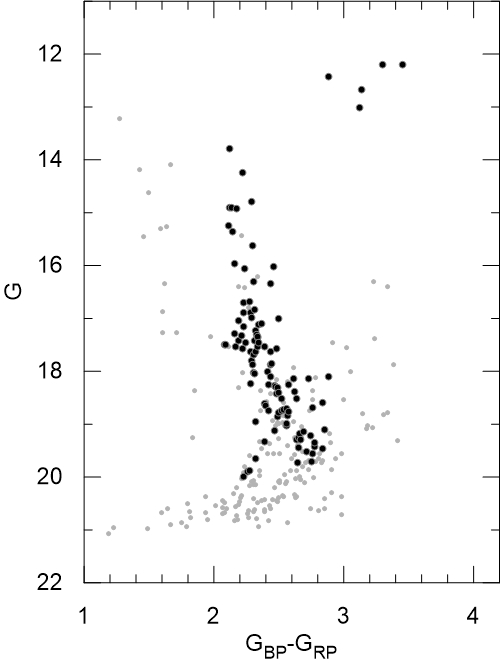}
	\caption {The $\mu_{\alpha}$ versus $\mu_{\delta}$ for 305 stars of Be~55 (filled dots). The field stars inside 15 arcmin are shown with small grey dots. There are probable 109 members in the red circle as the fitted proper motion circle (0.5 mas~yr$^{-1}$). Gaia EDR3 parallax ($\varpi$) histogram (left panel) for 305 (blue ) and 109 members (red). On $(G, G_{BP}-G_{RP})$, a single stellar cluster sequences of the probable members are clearly visible.}
\end{figure}

\begin{figure}[!t]\label{Fig-5}
	\begin{center}
		\includegraphics[width=6.5cm, height=11cm] {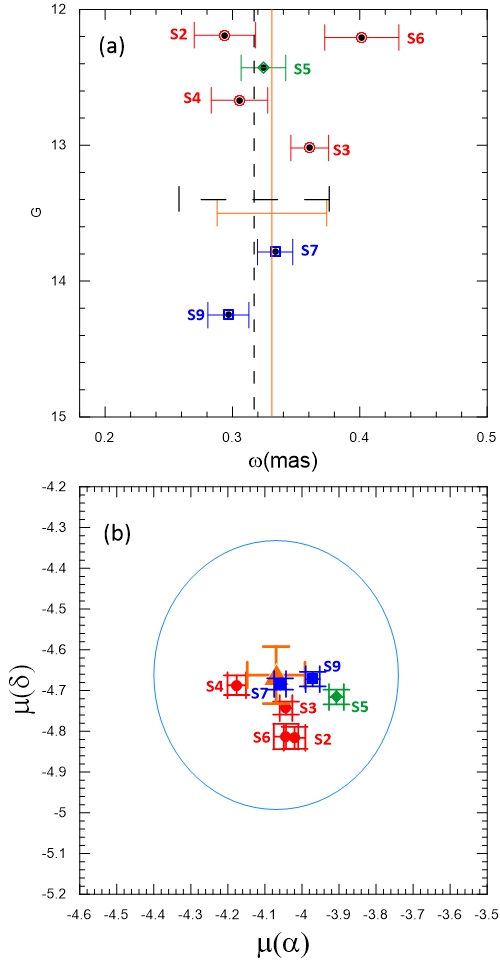}
		\caption{GaiaEDR3 parallax ($\varpi$) against G~mag (panel a) and the $\mu_{\alpha}$ versus $\mu_{\delta}$ (panel b) for five RSGs/RBGs and two BS candidates. The median Gaia EDR3/photometric parallaxes are shown with the vertical dashed black and solid orange lines together their uncertainties (horizontal lines), respectively. The orange triangle of panel (b) denotes the median Gaia EDR3 proper motion components of Be~55. The blue circle represents  the expected dispersion $(0.33~mas~yr^{-1})$ in proper motion.}
		\label{fig:CCVUB}
	\end{center}
\end{figure}

\subsection{Membership Selection}

As is seen in the papers of \cite{akk19} and \cite{akk20}, $UBVI$ photometric data of Be~55 have been combined with Gaia EDR3 proper motion and parallax data \citep{bro20,lin20} to classify the likely cluster members. On the ($\mu_{\alpha},\mu_{\delta}$) plane (Fig.~4), 305 stars (filled dots) within $R_{RDP}= 2'.5$ are displayed. The grey dots represent the field background/foreground stars with a region centered around 15 arcmin. The fitted radius (red circle) by eye is determined as 0.5 mas~yr$^{-1}$ by utilising the histograms (shown with red lines) attached to the $\mu_{\alpha}$ versus~$\mu_{\delta}$ plot. 109 probable members in the red circle are used to calculate from both the median  values ($<\mu_{\rm \alpha}>$ and $<\mu_{\rm \delta}>$) and the quantity $\mu_{\rm R}=\sqrt{ (\mu_{\rm\alpha}-<\mu_{\rm \alpha}>)^{2} + (\mu_{\rm \delta}-<\mu_{\rm \delta}>)^2}$). 
Parallax histogram and $(G, G_{BP}-G_{RP})$ diagram are presented in the bottom of Fig.~4. The blue and black histograms show 305 and 109 members, respectively. The number of 109 members is in good agreement with the 107 members, classified by \cite{can18,can20}. The applied median Gaia EDR3 parallax ($\varpi = 0.317\pm0.059$~ mas) on the histogram gives 64 members, and these provide a single stellar sequence (black dots) in the $(G, G_{BP}-G_{RP})$ diagram. Since the current Gaia EDR3 parallaxes of the individual stars have a random error, the ensemble median parallax is considered. Accordingly the parallax error is the median of the members. The median astrometric values of 64 members are ($\mu_{\alpha},\mu_{\delta}$)=(-4.070$\pm$0.078,~-4.662$\pm$0.070) mas~yr$^{-1}$ and $\varpi = 0.317\pm0.059$~mas, respectively which are compatible with ($\mu_{\alpha},\mu_{\delta}$)=(-4.050$\pm$0.219,-4.618$\pm$0.214) mas~yr$^{-1}$ and $\varpi = 0.309\pm0.091$~mas, given by \cite{can18,can20}.

The determination of the age of Be~55 depends on the memberships of five RSGs/RBGs and two BSs. As was done by L18 (see their figs.~7-10), to querry on their memberships further, their Gaia EDR3 parallaxes against G-mag plus $\mu_{\alpha}$ versus $\mu_{\delta}$ are displayed in Fig.~5(a) and (b). In panel (a) the vertical lines indicate the median Gaia EDR3 parallax  $\varpi$ = 0.317$\pm$0.059 mas (dashed black line), and photometric parallax $\varpi$ = 0.331$\pm$0.043 from $Q_{V\lambda}$--$Q'$ (orange line, see section 3.4), respectively.  As is noticed from panel (a), the parallax uncertainties of five RSGs/RBGs and two BSs (Table 5) almost remain within the error limits of the vertical/horizontal lines. Note that $S6$/$L145$'s parallax value is far from the median parallax. But its uncertainty remains the error limit of horizontal dashed line. Here, the notation "L" represents the star ID in L18.

The proper motions of five RSGs/RBGs and two BSs together their errors are inside the borders of the blue circle (panel b). Blue circle is constructed as the following, the derivation of the relation $V_{tan}=4.74\mu/\varpi$ provides $\sigma_{V(tan)} = 4.74 \sigma_{\mu}/\varpi$. In this case, it is assumed that the adopted $\varpi$ has no error, and that the velocity of stars in Be~55 is homogeneous and isotropic, $\sigma_{V(tan)} = \sigma_{V(rad)}$. A20 give the average radial velocity of the bright giants of Be~55 as $V_{rad} = -27.7\pm4.9$~km s$^{-1}$.  Then the expected dispersion in proper motion is obtained as $\sigma_{\mu} = (4.9~km s^{-1})\times\varpi (=0.317~mas) /4.74$ = 0.33~mas~yr$^{-1}$. 
In that case these all seem to be members, for this respect the derived age in this paper is safely interpreted. The global parallax zero point shift, $\Delta\varpi=-0.017~mas$ of Gaia EDR3 \citep{bro20} is applied to the median value, $0.317$~mas, it gives a close distance up to 0.16 pc.

The distribution of 64 probable members on the CMDs, $(V, U-B)$, $(V, B-V)$, and $(V, V-I)$ is presented in Fig.~6.  Maidanak $UBVI~CCD$ observations of the cluster on the CMDs are shown with grey symbols; 89 (panel a), 357 (panel b), 403 (panel c), respectively.

\begin{figure}[!t]
	\begin{center}
		\includegraphics[width=7cm, height=16cm] {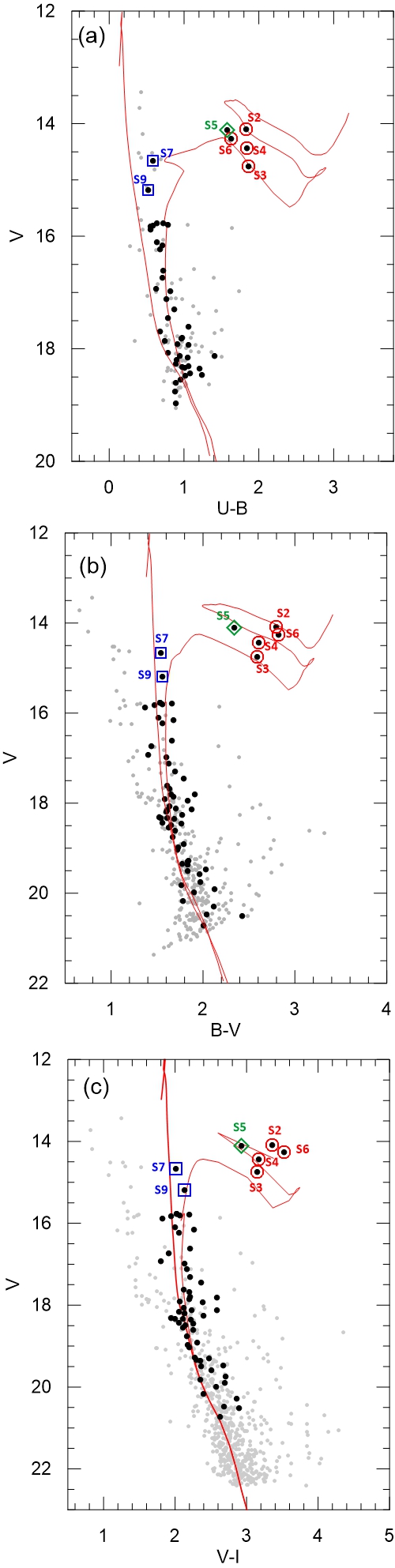}
		\caption{The CMDs of the members; $V$-$(U-B)$ (N$=$46, panel~a), $V$-$(B-V)$ (N$=$64, panel~b) and $V$-$(V-I)$ (N$=$64, panel~c). Solid red lines/curves denote the E12 ZAMS/85 Myr isochrones. Maidanak $UBVI~CCD$ observations are shown with grey symbols; 89 (panel a), 357 (panel b), 403 (panel c), respectively. Diamond, filled red circles, and blue squares represent one Cepheid, four RSGs/RBGs and two BSs, respectively. Star designation is from N12.}
	\end{center}
\end{figure}

\begin{figure}[!t]
	\begin{center}
		\includegraphics[width=8cm, height=9cm] {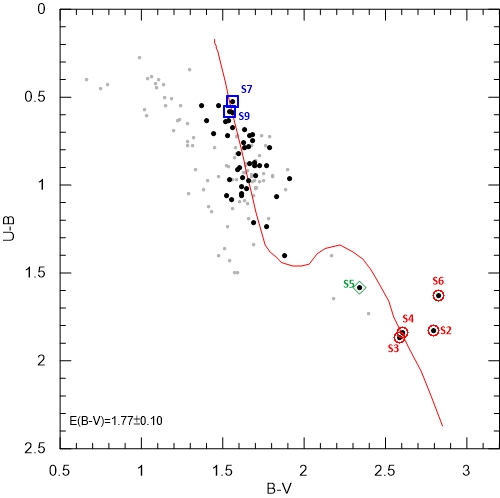}
		\caption{$(U-B), (B-V)$ (CC) diagram for 38 members (filled dots) of Be 55. Grey symbols denote the 97 stars from Maidanak $UBV~CCD$ observations. Red curve shows the reddened SK82 main sequence. The symbols of bright evolved stars are the same as Fig.~6.}
		\label{fig:CCVUB}
	\end{center}
\end{figure}

\subsection{Reddening}

The reddening of Be~55 is determined from 38 early type members with $V < 16$~mag and $(U-B) < 1.1$ on the $(U-B),(B-V)$ (CC) colour-colour plot (Fig.~7). Out of 64 members, 46 members have UBV data.  Five members with a red circle are the RSGs/RBGs. The mean reddening from these early type members is estimated as $E(B-V)$=1.77$\pm$0.10~mag, and  thus the reddened colour sequence of the Schmidt-Kaler (SK82) \citep{sch82} (red curve) is fitted to the CC diagram. For the determination of $E(B-V)$, the reddenings $E(V-I)$, $E(V-J)$, $E(V-H)$, and $E(V-K_{S})$ have been estimated by using the intrinsic colour relation of early type stars, given by S13 (see their tables~2--3). Here the colour excess ratio $E(U-B)=0.72E(B-V)+0.025E(B-V)^2$ of S13 is adopted. The reddening law of Be~55 using colour excess ratios $E(V-\lambda)$ for IJHK$_{S}$ photometry has been tested in Fig.~8. 38 members (filled blue dots) lie on the solid line. The total-to-selective extinction ratio is obtained as $R_{V}=3.13\pm0.04$ (Table 4), which implies that the reddening law toward Be~55 is quite normal. According to \cite{gue89}, colour excess ratio of optical-near infrared colours is related to the total-to-selective extinction ratio.

Its highly reddened value, $E(B-V)=1.77\pm0.10$~mag is in agreement with $E(B-V)=1.85$~mag of N12, $E(B-V)=1.81\pm0.15$~mag of A20, $E(B-V)=1.75$~mag of \cite{mol18}, $E(B-V)=1.74$~mag of L18, and $E(B-V)=1.74\pm0.10$~mag of \cite{mac07} within the uncertainties (Table 7), respectively. This reddening is larger than the ones of \cite{tad08} and \cite{buk11}. Note that N12 also give $E(J-K_{S})=0.85$~mag, which converts to $E(B-V)=1.73$~mag via a relation $E(J-K_{S})=0.49E(B-V)$ of \cite{dut02}. 

The reddening value derived from the early type stars cannot be used to deredden late-type supergiants. Therefore, 
the mean reddening value from their $(J-K_{s})$ colours of the five evolved stars (Table 5) is estimated as $E(J-K_{s})=0.92\pm0.06$~mag, by utilising the intrinsic colour, $(J-K_{s})_{0}$ of \cite{koo83}. This converts to $E(B-V)=1.84\pm0.11$~mag from the relation $E(J-K_{s})=0.49 E(B-V)$ \citep{dut02}. Their reddenings are listed in Col.~5 of Table~5.
For Be~55, A20 mention a non-negligible differential reddening. The reddened values $E(B-V)=1.77\pm0.10$~mag (early type stars) and $E(B-V)=1.84\pm0.11$~mag (five evolved stars) imply a differential reddening in Be 55. The members selected in Fig.~7 are scattered over more than 0.3 mag in $(B-V)$ for a given $(U-B)$. Likewise, the members are scattered over more than 0.5 mag in $(G, G_{BP}-G_{RP})$ (Fig.~4). 

The spatial variation of reddening of Be~55 as the reddening map, is derived from its member stars (red dots) and field stars (black dots), and is presented in Fig.~9. The size of the dots is scaled to the magnitude of the star. The coloured line types  as iso-reddening contours represent different amounts of reddening $E(B-V)$. From Fig.~9, there seems to be a differential reddening across the cluster due to a slight variation of $E(B-V)$. Most young OCs with ages younger than 10 Myr show a differential reddening across the field of view. But for slightly old OCs, it is not well known whether there is a differential reddening across the cluster or not. That's the reason to check the differential reddening across the cluster.

\renewcommand{\arraystretch}{1.1}
\begin{table}[!t]\label{Tables-04}
	\begin{center} 
		\caption {E(V-$\lambda$)/E(B-V) ratios (Col.~2) in terms of four colour indices (Col.~1). R$_{V}$ is the weighted average of four colours. Here $\lambda$ is I, J, H and K$_{s}$. N (last column): cluster star numbers.}
		\setlength{\tabcolsep}{0.26cm}
		\begin{tabular}{lll}
			\hline
			Colour & E(V-$\lambda$)/E(B-V)  & N \\
			\hline
			\
			V-I      &1.300$\pm$0.050&38  \\
			V-J      &2.341$\pm$0.131&38  \\
			V-H      &2.659$\pm$0.172&38 \\
			V-K$_{s}$&2.841$\pm$0.177&38 \\[2mm]
			&   R$_{V}$=3.13$\pm$0.04    \\
			\hline
		\end{tabular} 
	\end{center} 
\end{table}

\begin{figure}[!t]
	\begin{center}
		\includegraphics[width=8cm, height=8cm] {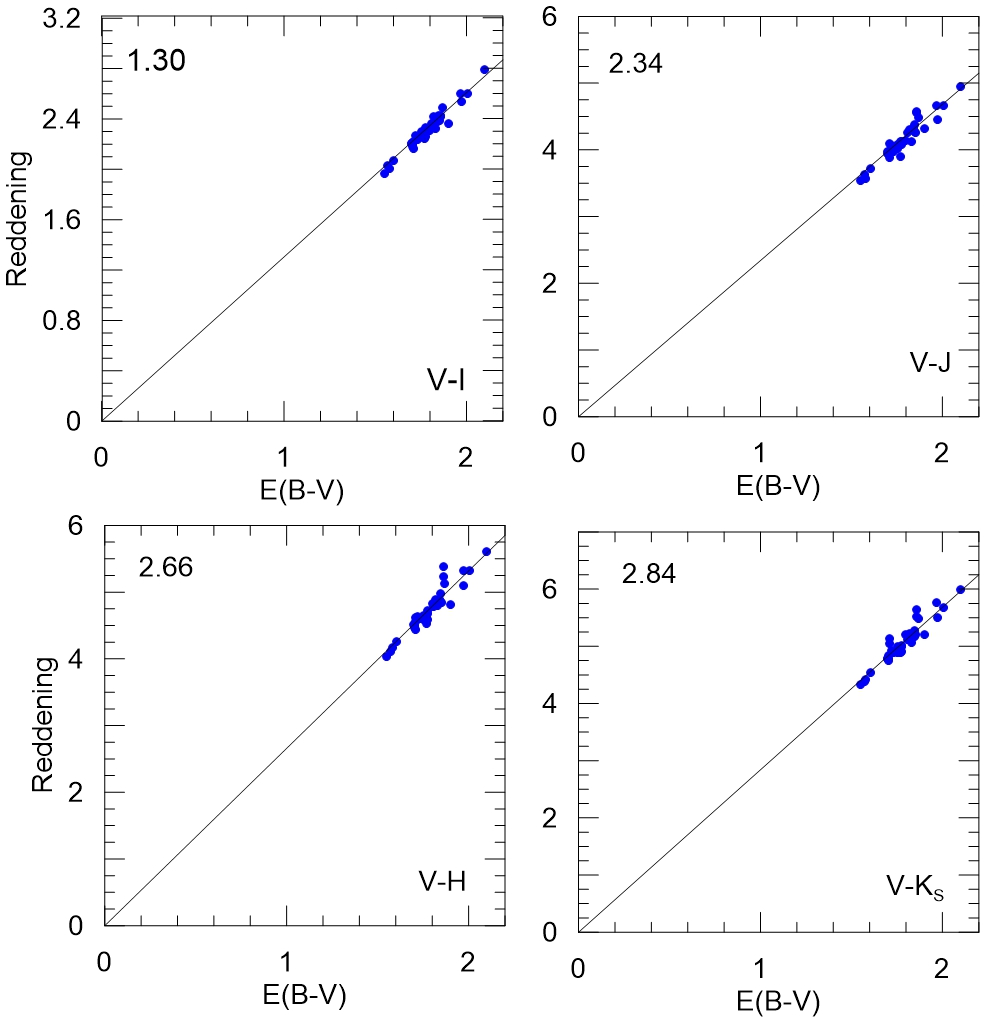}
		\caption{$E(V-\lambda)$ versus E(B-V) relations. $\lambda$ denotes IJHK$_{S}$ photometry.The solid line means $R_V = 3.1$. The colour excess ratios from the $IJK_{S}$ data consistently show that the reddening law toward Be~55 is normal.}
		\label{fig:Reddening}
	\end{center}
\end{figure}

\begin{figure}\label{F-figures-03}
	\centering{\includegraphics[width=0.9\columnwidth]{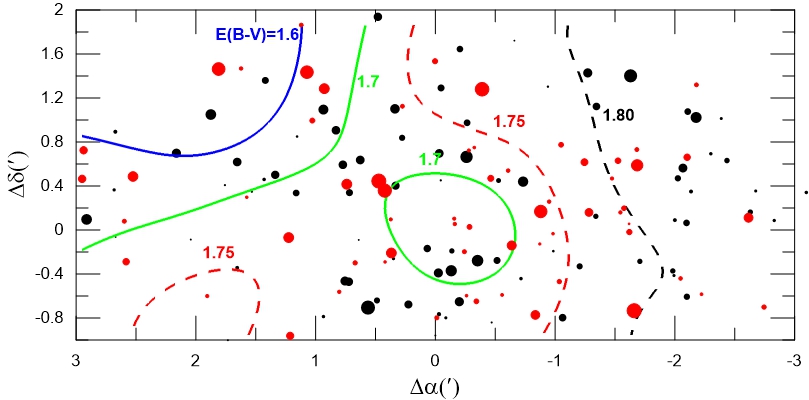}}
	\caption {Reddening map of Be~55. Red and black dots shows the member and the field stars, respectively. The size of the dots is scaled to the magnitude of the star. 
The coloured line types as iso-reddening contours represent different amounts of reddening $E(B-V)$.}
\end{figure}

\subsection{Distance}

Instead of the reddening-corrected CMDs, distance modulus/distance of Be~55 is obtained by using the ZAMS fitting method with the reddening-independent indices (S13). For this, UBVRIJHK$_{s}$ photometry of 38 early type stars is used.  The reddening-independent quantities, $Q'$, $Q_{V\!I}$, $Q_{V\!J}$, $Q_{V\!H}$ and $Q_{V\!K_{S}}$ are utilised. The $Q_{V\lambda}$--$Q'$ plots for 38 early type members have been displayed in Fig.~10.
Solid lines denote the fitted ZAMS relation of \cite{sun13} to the members. As suggested by \cite{lim14}, ZAMS should be fitted to the lower ridge of the MS band to avoid the effects of multiplicity and evolution. The ZAMS line is shifted up and down in Fig.~10 by 0.1 mag. The error of this method is about 0.20 mag. Once the ZAMS has been adjusted above and below the distribution of the members (Fig.~10),  the distance modulus from four colour indices is obtained as $(V_{0}-M_{V})=12.40\pm0.20$~mag, equivalent to 3.02$\pm$0.28~kpc, which are adopted for this paper. The median Gaia EDR3 parallax is $\varpi$ = 0.317$\pm$0.059 mas, which corresponds to  $(V_{0}-M_{V}) = 12.49\pm0.41$~mag (d$=$3.15$\pm$0.59 kpc). The two distances are compatible within the uncertainties. 

\begin{figure}[!t]
	\begin{center}
		\includegraphics[width=8cm, height=8cm] {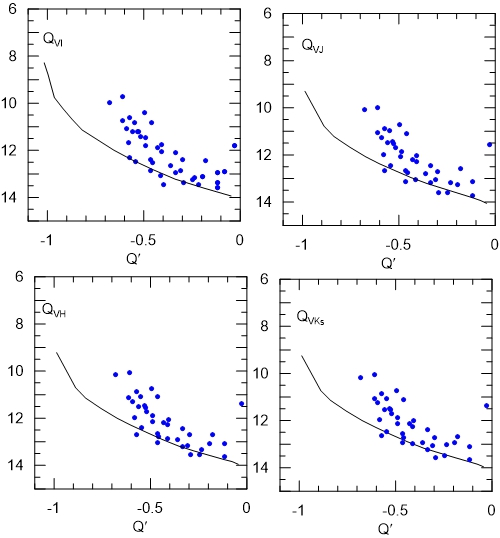}
		\caption{$Q_{V\lambda}$ versus $Q'$ diagrams (38 early type members, filled blue dots) for determination of distance modulus of Be~55.  $\lambda$ denotes IJHK$_{S}$ filters. A careful ZAMS fitting to the lower boundary of the MS band was carried out in the reddening-independent $Q_{V\lambda}$-$Q^{\prime}$ planes. The ZAMS relation of S13 is used to determine the distance to the cluster after adjusting by $12.4 \pm 0.2$, respectively.}
	\end{center}
\end{figure}

\renewcommand{\arraystretch}{1.2}
\begin{table*}
	\centering
	\setlength{\tabcolsep}{0.1cm}
	\caption{Properties of seven bright stars in Be~55.}
	\begin{tabular}{lcccccccccr}
		\hline
		Star no & V & (B-V)&(U-B)&E(J-K$_{S}$)/E(B-V) &$m~(M_{\odot})$&M$_{V}$  &$\varpi$&$d_{Gaia}$ &$d_{ph.}$  &SpT   \\
		(N12/L18) &mag &  & & & &mag &mas &kpc &kpc  & (A20)  \\
		\hline
		$S5$/$L107$&14.106 &2.340&1.582&0.863/1.726&5.9&-3.99& 0.324$\pm$0.017 &3.09$\pm$0.16&3.47$\pm$0.45& F8Ib  \\
		
		$S3$/$L163$&14.754 &2.588&1.866&0.933/1.866&5.8&-3.35 &0.360$\pm$0.015&2.78$\pm$0.12&2.84$\pm$0.37& G8II   \\
		
		$S4$/$L110$&14.437 &2.604&1.840&0.885/1.770&5.9&-3.67  & 0.305$\pm$0.022&3.28$\pm$0.24&3.26$\pm$0.42& K0Ib-II \\
		
		$S2$/$L196$&14.094 &2.796&1.826&1.006/2.012& 6.0&-4.01 &0.294$\pm$0.024 &3.40$\pm$0.28&2.30$\pm$0.30& K0Ib \\
		
		$S6$/$L145$&14.266 &2.825&1.628&0.906/1.812& 6.0 &-3.84 & 0.402$\pm$0.029&2.49$\pm$0.18&3.07$\pm$0.40& K4II \\
		
		$S9$/$L198$&15.186 &1.562&0.524&-/1.770 &7.5&-2.70&0.297$\pm$0.016&3.37$\pm$0.18&2.95$\pm$0.38&  B3-4IIIShell  \\
		
		$S7$/$L94$& 14.666 & 1.543&0.581&-/1.770 &7.5&-3.22&0.334$\pm$0.014&2.99$\pm$0.13&2.95$\pm$0.38&B4IV \\
		\hline
	\end{tabular}  
\end{table*}

\section{Age of Be~55}

$M_{bol}-\log T_{eff}$ diagram (HRD) for 46 members (41 early type and five RSGs/RBGs) is presented in Fig.~11. 
Their effective temperature and $BC$ values have been determined from their $(U-B)_{0}$ and $(B-V)_{0}$ by utilising table 5 of S13. Their $M_{bol}$ values are obtained from the relation $M_{bol} =BC+V-R_{V}E(B-V)-(V_{0}-M_{V})$. For this, the UBV photometry of 41 early type and five RSGs/RBGs is de-reddened by $E(B-V)=1.77$~mag and $E(B-V)=1.84$~mag, respectively.

The metal abundance, $[M/H]=0.07\pm0.12$ ($Z = 0.014$) of Be~55, given by L18  is considered for selecting the isochrone. Note that Be~55 has a extended main sequence. Due to the diversity of the rotational velocity even in a cluster \citep{lim19}, MS turn-off broadens. Broadening in the main sequence of the CMD could be also due to binarity, peculiarity, rotation in the stars. Therefore, the solar abundance Geneva models of \cite{eks12} (hereafter E12) ($V/V_{crit} = 0.4$) and \cite{geo13} (hereafter G13) (for B-type stars and $V/V_{crit} = 0.3$)  have been fitted to the members in the $M_{bol}-\log T_{eff}$ (Fig.~11) and de-reddened $(V_{0}, B-V_{0})$ (Fig.~12). 
$(V_{0}-M_{V})=12.40\pm0.20$~mag (Q technique) and the reddenings  $E(B-V)$=1.77$\pm$0.10~mag (for early type stars) and  $E(B-V)=1.84\pm0.11$~mag (for evolved stars) have been applied to the E12 and G13 isochrones while fitting.

As is seen in Fig.~11, the 85$\pm13$ Myr E12 isochrone with high rotation (blue curve) describes general morphology of HRD without accurate locations. The 85 Myr E12 isochrone (blue curve) is almost consistent with the RSGs/RBGs, $S4$/$L110$, $S5$/$L107$ (Cepheid), and $S6$/$L145$, except for $S3$/$L163$. Also, the 70$\pm10$ Myr G13 isochrone (red curve) with moderate rotation for B-type stars provides good fit with the RSGs/RBGs, $S2$/$L196$, $S4$/$L110$ and $S5$/$L107$ (Cepheid), except for $S3$/$L163$ and $S6$/$L145$. Note that $S5$/$L107$ appears to be on the core-He burning blue loop.  Note that non-rotating E12 isochrone (blue dashed line) gives a faint turn-off, and going below the RSGs/RBGs.

On the $V_{0}-(B-V)_{0}$ (Fig.~12), $S2$/$L196$ and three RSGs/RBGs occupy positions which are consistent with the 85 Myr E12 isochrone. However, the 70 Myr G13 isochrone is not compatible with the positions of the RSGs/RBGs.
	
The two BSs, $S9$/$L198$ and $S7$/$L94$ locate between ZAMS and the 45 Myr isochrone (Figs.~11-12). The age of BSs seems to be definitely younger than the cluster. 
From Figs.~11-12, there appears an extended main-sequence turn-off (eMSTO), since the main sequence band is expected to broaden in CMDs due to their different main sequence lifetimes. 
Also, the broadening in the MS band may also be due to differential reddening across the cluster because the scatter in $E(B-V)$ is about 0.1 mag (section 3.3). One can expect about 0.3 mag scatter $(3\sigma)$. since the width of MS band is about 0.3 mag.

\begin{figure}[!t]
	\begin{center}
		\includegraphics[width=8cm, height=8cm]{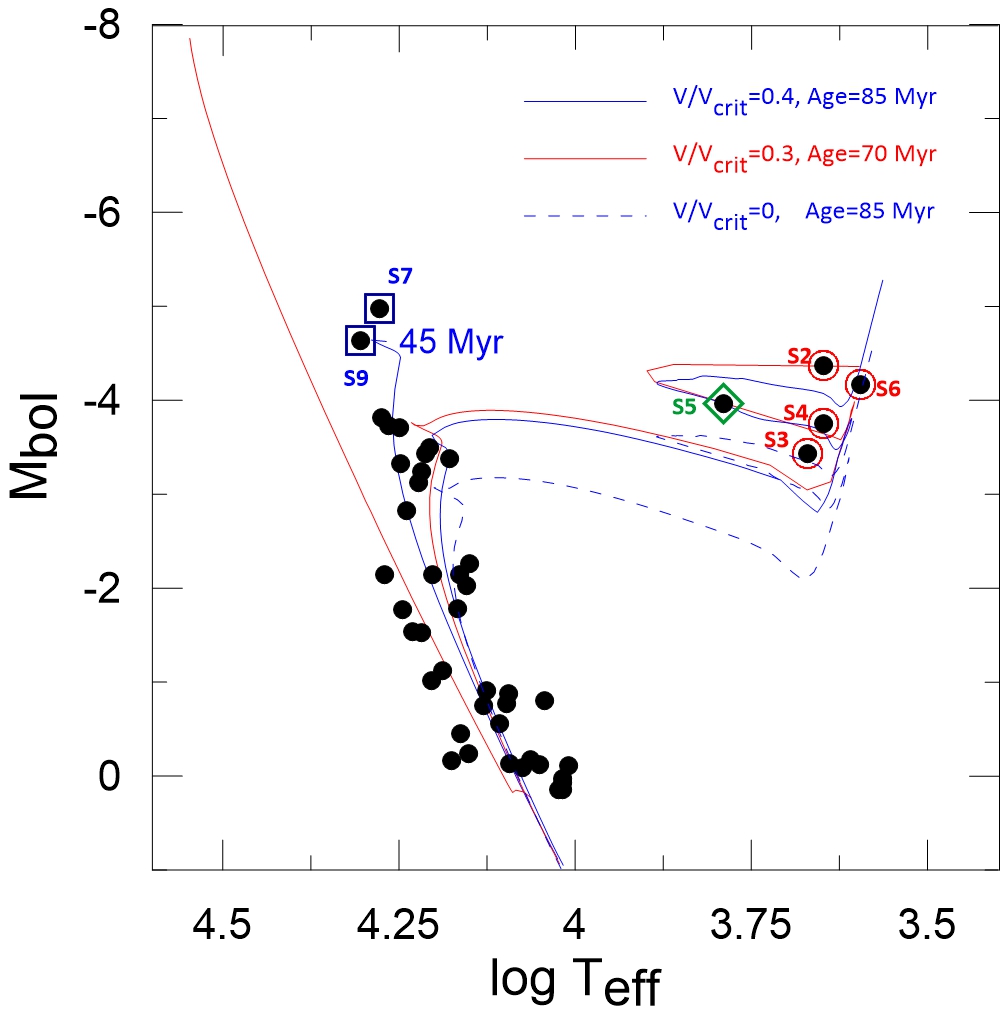} \hspace*{1mm}
		\caption{$M_{bol}$ versus $\log T_{eff}$ of 46 probable members.  Solid and dashed blue curves represent rotating/non-rotating E12 isochrones. Red curve shows G13 isochrone for B type stars with $V/V_{crit} = 0.3$. The solid red line represents the ZAMS of E12. The symbols are the same as Fig.~6.}
	\end{center}
\end{figure}

\begin{figure}[!t]
	\begin{center}
		\includegraphics[width=8cm, height=8cm] {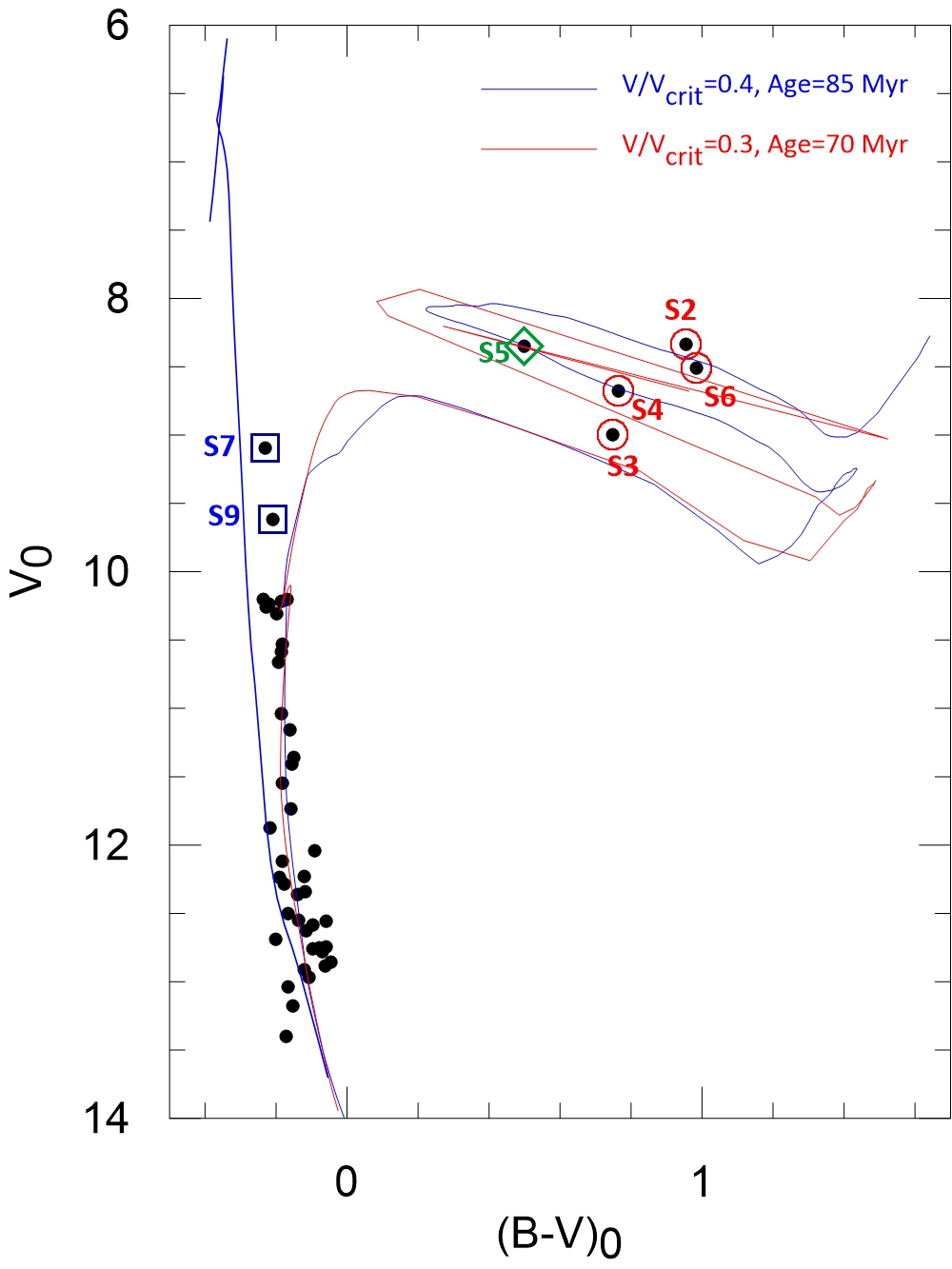}
		\caption{$V_{0}$-$(B-V)_{0}$ of 46 probable members. The symbols are the same as Fig.~11.}
	\end{center}
\end{figure}

\begin{figure}[!t]\label{Fig-11}
	\begin {center}
	\includegraphics[width=6cm, height=10cm] {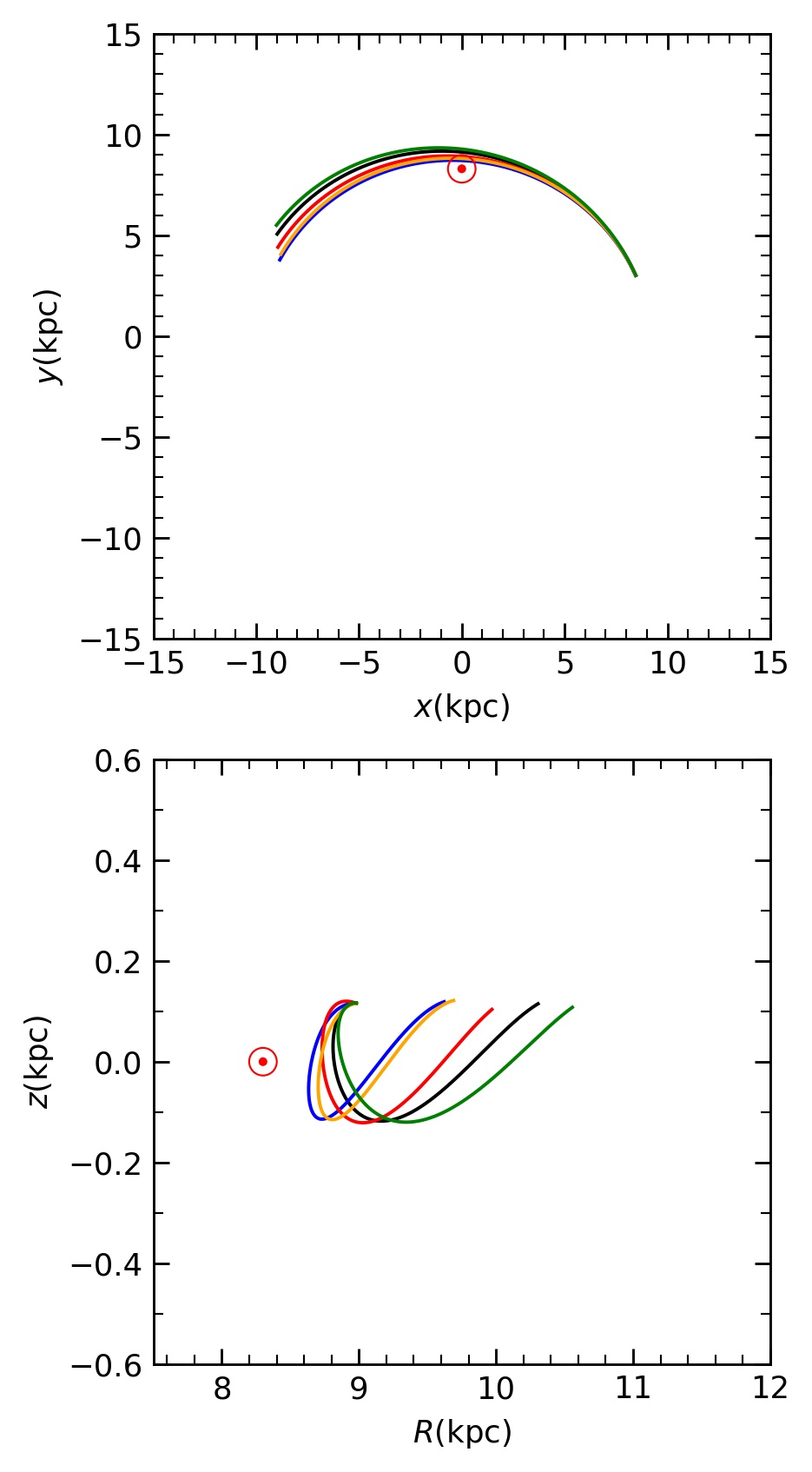}
	\caption{The orbits of five RSGs/RBGs on x--y (kpc) (top panel)  and z--R (kpc) (bottom panel). Large circle shows the position of the Sun, $(z_{\odot},~R_{\odot})=(0,~8.2~kpc)$.}
	\end {center}
\end{figure}

\section{Kinematics and orbital parameters}

From the radial velocities of A20 and the Gaia EDR3 proper motions for  five RSGs/RBGs, 
their heliocentric velocities ($U$, $V$, $W$) in the right-hand system  have been calculated from the algorithm of \cite{joh87}. The photometric distance of Be~55 ($3.02\pm0.28$~kpc) is adopted, instead of their individual parallaxes. 
These space velocities are transformed to the components $U'$, $V'$, $W'$ by correcting for the Solar motion $(U, V, W)_{\odot} = (+11.10, +12.24, +7.25)$ km s$^{-1}$ with respect to the local standard of rest (LSR) \cite{sch10}. Here, R$_{\odot}$$=$8.2$\pm$0.1 kpc \citep{bg16} and $V_{LSR}$ = 239\,km\,s$^{-1}$ \citep{bru11} are adopted. The heliocentric cartesian distances ($x'$, $y'$, $z'$) (kpc) and LRS-velocity components ($U'$, $V'$, $W'$) have been transformed to Galactic Rest of Frame (GSR) i.e., ($x$, $y$, $z$) (kpc) and ($V_{x}$, $V_{y}$, $V_{z}$) from the equations of  \cite{kep07}. The Galactocentric velocity component ($V_{\Phi}$) (km s$^{-1}$) (or azimuthal velocity) in a cylindrical frame is estimated via $V_{\Phi} =  \frac{x V_{y} - y V_{x}}{R}$. Here, $V_{\Phi}<0$ means prograde.  By utilising the "MWPotential2014" code in the galpy-code library \footnote[1]{http://github.com/jobovy/galpy} written by \cite{bov15}, peri- and apo-galactic distances $(R_{min},~R_{max})$ (kpc) and the maximum height distance (z$_{max}$) (kpc) have been obtained. The orbital eccentricity (ecc) is estimated via the relation $e = (R_{max}-R_{min})/(R_{max}+R_{min})$. Five evolved member's orbits have been integrated for 85 Myr within the Galactic potential. The galactic potential as sum of the Galactic components is explained by \cite{bov15}. Their orbital angular momentum components $J_{x}$, $J_{y}$, $J_{z}$ and J$_{\bot}$ (kpc km s$^{-1}$) are calculated from the equations of \cite{kep07}. 
The total angular momentum J$_{\bot}$ is defined as $J_{\bot}=(J_{x}^2+J_{y}^2)^{1/2}$. The relations of x-y~(kpc) and z - R~(kpc) of the five RSGs/RBGs are shown in Fig.~13.

\renewcommand{\arraystretch}{1.2}
\begin{table}[!h]\label{k}
	\centering
	\tiny
	\caption{For the available Gaia EDR3 proper motion components (mas~yr$^{-1}$) and spectroscopic radial velocity data (km s$^{-1}$) of five RSGs/RBGs, kinematics ($U$, $V$, $W$, $V_{\Phi}$) km s$^{-1}$, orbital parameters (R$_{max}$, R$_{min}$, z$_{max}$)~(kpc), eccentricity (ecc), and angular momentums  ($J_{z}$ and $J_{\bot}$) (kpc km s$^{-1}$), respectively.}
	\setlength{\tabcolsep}{0.2cm}
	\begin{tabular}{ccccc}
		\hline
		Star &$\mu_{\alpha}$  &$\mu_{\delta}$ & $V_{rad}$& \\
		\hline
		S2 &-4.020$\pm$0.029&-4.816$\pm$0.027&-24.55$\pm$0.39\\
		S3 &-4.043$\pm$0.017&-4.743$\pm$0.016&-33.33$\pm$0.47\\
		S4 &-4.176$\pm$0.025&-4.687$\pm$0.024&-28.97$\pm$0.47\\
		S5 &-3.907$\pm$0.020&-4.716$\pm$0.018&-31.63$\pm$0.86\\
		S6 &-4.043$\pm$0.033&-4.813$\pm$0.031&-21.61$\pm$0.60\\
		&            &            &        &        \\
		\hline
		& U     & V      & W     & $V_{\Phi}$ \\
		\hline
		S2 & 90.71 & -19.59 & -7.55 & -252.38 \\	
		S3 & 90.66 & -28.39 & -6.88 & -244.08\\
		S4 & 91.18 & -24.04 & -4.84 & -248.35 \\	
		S5 & 88.94 & -26.72 & -7.97 & -245.07\\
		S6 & 90.76 & -16.60 & -7.23 & -255.22\\
		&        &        &       &          \\
		\hline
		&R$_{min}$ & R$_{max}$ &   ecc   \\
		\hline
		S2 &  8.81 &10.31  &0.08 \\
		S3 &  8.63 & 9.62  & 0.05 \\
		S4 &  8.73 & 9.97  & 0.07 \\
		S5 &  8.70 & 9.69  & 0.05 \\
		S6 &  8.85 &10.56  & 0.09 \\
		&        &       &        \\
		\hline
		& z$_{max}$ & $J_{z}$ & $J_{\bot}$  &     \\
		\hline
		S2 & 0.12 &-2266 & 24    \\
		S3 & 0.12 &-2192 & 24    \\
		S4 & 0.12 &-2230 & 33   \\
		S5 & 0.12 &-2200 & 24   \\
		S6 & 0.13 &-2292 & 24   \\
		\hline
	\end{tabular}  
\end{table}

\renewcommand{\arraystretch}{1.2}
\begin{table*}[!h]\label{Literature}
	\centering
	\tiny
	\caption{Comparison with the literature for Be~55.}
	\setlength{\tabcolsep}{0.1cm}
	\resizebox{\textwidth}{!}{
		\begin{tabular}{lcclllllll}
			\hline 
			&E(B-V) &$(V_{0}-M_{V})$& $d$~(kpc) & $Z$   &$\log~Age$ & Age& Isochrone & Photometry&   Ref.\\
			&mag & mag &kpc &  & & Myr& & &\\
			\hline 
						&1.77$\pm$0.10 &12.40$\pm$0.20 &3.02$\pm$0.28 &0.014&7.93$\pm$0.06&85$\pm$13&\cite{eks12}&CCD~$U\!BV\!I$ &This paper \\
			&1.74$\pm$0.07 &11.71$\pm$0.30 &2.20$\pm$0.30 & &7.80&63$\pm$12& PLR of S5 Cepheid &CCD~$VR$ &\cite{loh18} \\
			&1.85$\pm$0.16  &13.0$\pm$0.30 &3.98$\pm$0.55&solar&7.70&50$\pm$10&\cite{mar08}&CCD~$UBV$ &\cite{neg12} \\
			&1.81$\pm$0.15  &12.55$\pm$0.15 &3.24$\pm$0.22&solar&7.80$\pm$0.10&63$\pm$15&\cite{eks12}&CCD~$UBV$ &\cite{alo20} \\
			&1.75  &12.40 &3.02&solar&7.50-8.50&30-100&\cite{bre12}&$ugr$&\cite{mol18} \\
			&1.74$\pm$0.10  &10.42 &1.21$\pm$0.31 &solar&8.50&315&\cite{ber94} &CCD~$BV$ &\cite{mac07} \\
			&1.50  & &1.44$\pm$0.07&solar&8.48&300&\cite{bon04}& $2MASS-JHK_{s}$&\cite{tad08}\\ 
			&1.15  &12.23 &1.70$\pm$0.13&0.019&8.95&891&\cite{gir02}&$2MASS-JK_{s}$ &\cite{buk11}\\
			&1.55($A_{V}=4.8$) &12.20 &2.75&0.015&8.30&200&\cite{bre12}&$Gaia~DR2$ &\cite{can18,can20}\\
			\hline
		\end{tabular}  
	}
\end{table*}

\section{Discussion and Conclusion}

The photometric distance modulus/distance of Be~55 as $(V_{0}-M_{V}) = 12.40\pm0.20$~mag ($3.02\pm0.28$~kpc) is better and well consistent with the median Gaia~EDR3 distance (3.15$\pm$0.59~kpc). These distances locate Be~55 near the Perseus Spiral arm. 
The photometric distance of this paper is in concordance with $3.24\pm0.22$~kpc of A20 and 3.02~kpc of \cite{mol18} (Table~7). It is rather smaller than N12 but is farther than L18. L18'distance is from period-luminosity relation (PLR) of Cepheid $S5$ (Table~8). N12 obtain its distance modulus from the dereddened ZAMS of \cite{mer81} and SK82 as 13.0$\pm$0.30~mag on $M_{V}-(B-V)$. This corresponds to a distance, 3.98$\pm$0.55 kpc. L18'distance locates Be~55 on the outer edge of Local arm rather than in the Perseus arm as N12 suggest. A mismatch between the distances of L18 and N12 is explained by the underestimated uncertainties in N12's distance modulus. The other literature give close distances.

The distance 2.78 kpc of Cepheid $S5$/$L107$, from the PLR of \cite{laz20} is consistent with the ones of Gaia EDR3 and A20 within the errors (Table 8). Note that L18 give close distances. $M_{V}=-3.99$ from E12 isochrone estimates a large distance, 3.47~kpc.

\renewcommand{\arraystretch}{1.5}
\begin{table}[!t]\label{k}
	\centering
	\tiny
	\caption{Distances for the values of $V$, $M_{V}$, $P=5.85$~day, and $E(B-V)$ of Cepheid $S5$. The methods and their references for $M_{V}$ are listed in Cols.5--6.}
	\setlength{\tabcolsep}{0.1cm}
	\begin{tabular}{ccccccc}
		\hline
		d & V & $M_{V}$ & $E(B-V)$ & Methods for $M_{V}$ & Ref. & Remarks\\
		kpc & mag & mag & mag &   & & \\
		\hline
		3.47$\pm$0.45 & 14.106 & $-$3.99 &1.73 & E12 isochrone & 1 & This paper \\
		2.78$\pm$0.32 & 14.106 & $-$3.52 &1.73& $-$(2.67$\pm$0.16)(logP -1)-(4.14$\pm$0.05) &2 & This paper \\
		3.09$\pm$0.16 &   &     &       &      &  & Gaia EDR3\\
		2.20$\pm$0.30 & 13.834 & $-$3.23  &1.74 & $-$(2.88$\pm$0.18)logP-(1.02$\pm$0.16) & 3  & L18 \\
		2.40$\pm$0.30 &  13.834 & $-$3.93 &1.74 & $-$(2.43$\pm$0.12)(logP-1)-(4.5$\pm$0.02) & 4   &  L18 \\
		3.03$\pm$0.37 & 13.834 &   &    1.81 & a(logP-1)+b,~for JHK &  5    &    A20 \\
		\hline	
	\end{tabular}
	\begin{list}{Table Notes.}
		\item $[1]$:\cite{eks12}, $[2]$: \cite{laz20}, $[3]$: \cite{and13},	$[4]$:\cite{ben07}, $[5]$:\cite{che17}, 
	\end{list}
\end{table}

The E12 isochrone (high rotation) fittings to HRD/CMD derive turn-off age, 85$\pm$13 Myr of Be~55, by taking care five RSGs/RBGs.  For this age, the masses of five RSGs/RBGs from the E12 isochrone are about 6~$M_{\odot}$ (Col.~6 of Table 5). The 70 Myr G13 isochrone with moderate rotation does not provide a good fit to the RSGs/RBGs on the $V_{0}$-$(B-V)_{0}$ (Fig.12). The age 85 Myr is somewhat older than N12, L18, and A20 (Table~7) but falls in the range of 30--100 Myr of \cite{mol18}. From the non-rotating PARSEC isochrones, A20 give its age as $63\pm15$ Myr (see their fig.~6 and table 1), by considering evolved members. The age 85 Myr also falls in the range of 63--105 Myr, which is found from the period-age relations in table 14 of A20 for Cepheid $S5$/$\#107$. Note that N12 apply two isochrones of \cite{mar08}; $\log t=7.6$~(40~Myr) and $\log t=7.7$~ (50~Myr) on $(V, U-B)$. By taking $E(J-K_{S})=0.85$~mag, N12 also apply the same isochrones to  $(K_{S},~J-K_{S})$, and give an age 50 Myr. The other literature values find more old ages (Table~7), than this paper. 

From Table 7, Note that \cite{can18,can20} from Gaia DR2 photometry give a less reddening, a close distance and an old age, 200 Myr, as compared to the values of this paper.

The spread of the brightness of the RSGs/RBGs in the HRD/CMDs indicates a much large spread in age. This feature may be the result of the diversity of stellar rotation among evolved cluster members \citep{lim19,lim16,sun97}, which indicate the necessity of cluster isochrone with non-single rotational velocity distribution. The stars in OCs do not have the same rotational velocity \citep{lim19}. Some stars have very low rotational velocities, and some may be fast rotators.
For that case, even the same mass stars are not the same position in the HRD because fast rotators have longer MS lifetime. This feature also indicates the necessity of cluster isochrones with non-single rotational velocity distribution. As a result, the possible inconsistences on the locations of the RSGs/RBGs to the rotating/non-rotating isochrones may be resulting from the age spread of stars in young OCs. According to A20, the possible inconsistence of the locations of the RSGs/RBGs on HRD/CMD is due to the strong reddening, rather than metal abundance.

Two BS candidates in $V_{0}$-$(B-V)_{0}$ (Fig.~12) lie on $(V_{0} < 10.01~mag)$ ($V < 15.50$~mag) and  $((B-V)_{0} < 0.03)$ $(B-V) < 1.80$. These limits for $V_{0}$ and $(B$--$V)_{0}$ which BSs occupy in CMDs are similar to those given by \citet[][fig.~19]{car01} and \citet[][fig.~10]{car10}.  
The positions of two BSs locate between ZAMS and the 85 Myr isochrone. The age of BSs is definitely younger than the cluster. Therefore the 45 Myr isochrone is drawn up to MS turn-off. 
As discussed by \cite{fer16}, BSs are commonly defined as stars brighter and bluer than the main-sequence (MS) turnoff in open/globular clusters. Therefore, their origin cannot be explained with normal single star evolution. Two main formation mechanisms are proposed: (1) mass transfer in binary systems \citep{mc64} possibly up to the complete coalescence of the two stars, and (2) stellar collisions \citep{hil76}. Both these processes can potentially bring new hydrogen into the core and therefore 'rejuvenate' a star to its MS stage \citep{lom02,che09}. 
According to \cite{san53} and \cite{tou97}, the increase in mass of a star makes it look younger than it is.

The kinematics, orbital and angular momentum values of the five evolved members (Table~6) indicate that Be~55 is a member of Galactic thin disk population, which is also consistent with its metallicity, $[M/H]=0.07\pm0.12$. With the circular orbits, ecc$=$ [0.05,~0.09],  Be 55 does not seem to be completed a tour around the center of the Galaxy (top panel of Fig.~13). They reach to z $\sim$ 0.13 kpc, and their birth places are at $\sim$ 9 kpc (bottom panel of Fig.~13). However, their orbits show that the cluster passed a part of its time at $R_{min}=8.63-8.85$ kpc.

The total mass for 64 early type members is obtained as $M_{tot}=224~M_{\odot}$ from 85 Myr E12 isochrone. From the photometric distance (3.02$\pm$0.28~kpc) and angular size $\theta\sim0.073$ deg (0.0013 in rad), the diameter of Be~55 is determined as 3.85 pc. In order to check its  stability, the maximum tangential velocity of the members within the radius of $\mu=0.5$ mas~yr$^{-1}$ is estimated as 7.2~km s$^{-1}$, via the relation, $V_{tan}=4.74\mu\times d(kpc)$. These values indicate that its virial mass is about $M_{vir}=25100~M_{\odot}$, which is far larger than the cluster mass, 224~$M_{\odot}$. Be~55 with its total mass, 224~$M_{\odot}$,  depending on  its location of $R_{GC} = 9.04$ kpc and $\ell=93^{\circ}.03$ is a survivor against internal and external perturbations related to, e.g. stellar evolution, mass segregation, spiral arms, and encounters with the disk and giant molecular clouds.

\section*{Acknowledgements}

I thank H.~Sung  for providing his private photometric  data, and the interpretations on the analysis. Y.~Karatas and H.~ Cakmak are also thanked for the kinematics and dynamics. The referee is thanked for the useful suggestions. This paper has made use of results from the European Space
Agency (ESA) space mission Gaia, the data from which were processed by the
Gaia Data Processing and Analysis Consortium (DPAC). Funding for the DPAC
has been provided by national institutions, in particular the institutions participating in the Gaia Multilateral Agreement. The Gaia mission website is http:
//www.cosmos.esa.int/gaia.


\vspace{-1em}

\begin{theunbibliography}{}
\vspace{-1.5em}

	
	
	
	\bibitem[Akkaya Oralhan et al.\ (2019)]{akk19} Akkaya Oralhan, \'I., Michel, R., Schuster, W.J, Karata\c{s}, Y., Karsl\i, Y., Chavarr\'ia-K, C., 2019, Journal of Astrophysics and Astronomy, 40,33
	
	
	\bibitem[Akkaya Oralhan et al.\ (2020)]{akk20} Akkaya Oralhan, \'I., Michel, R., Karsl\i, Y., Cakmak, H., Sung, H., Karata\c{s}, Y, 2020, Astronomische Nachritten, 341, 44
	
	\bibitem[Alonso-Santiago et al.\ (2017)]{alo17} Alonso-Santiago, J., Negueruela, I.,  Marco, A., Tabernero, H.M., Gonzalez-Fernandez,C.,  Castro, N., 2017, \mnras, 469, 1330

    \bibitem[Alonso-Santiago et al.\ (2020)]{alo20} Alonso-Santiago, J., Negueruela, I.,  Marco, A., Tabernero, H.M., and  Castro, N., 2020, arXiv:2009.124118v, \aap, ?,? (A20)
	
	\bibitem[Anderson et al.\ (2013)]{and13} Anderson, R.I., Eyer, L., Mowlavi, N., 2013, \mnras, 434, 2238
	
	\bibitem[Arellano Ferro et al. \ (2003)]{are03} Arellano Ferro, A., Giridhar, S., Rojo, Arellano, E., 2003,  RevMexAA, 39, 3
	
	\bibitem[Bastian and de Mink \ (2009)]{bas09} Bastian N., de Mink S. E., 2009, \mnras, 398, L11
	
	
	\bibitem[Benedict et al.\ (2007)]{ben07} Benedict, G.F. et al., 2007, \aj, 133, 1810
	
	\bibitem[\protect\citeauthoryear{Bertelli}{1994}]{ber94} 
	Bertelli, G., Bressan, A., Chiosi, C., Fagotto, F., \& Nasi, E., 1994, A\&AS , 106, 275
	
	\bibitem[Bland-Hawthorn et al.\ (2016)]{bg16} Bland-Hawthorn, J. and Ortwin Gerhard, O. 2016, \araa, 54, 529
	
	\bibitem[\protect\citeauthoryear{Bonatto et al.}{2004}]{bon04}
	Bonatto Ch., Bica E., Girardi L., 2004, \aap, 415, 571
	
	\bibitem[Bonatto and Bica \ (2007)]{bon07} Bonatto, Ch., Bica, E., 2007, \aap, 473, 445
	
	
	\bibitem[Bovy (2015)]{bov15} Bovy, J., 2015, ApJS, 216, 29
	
	
	\bibitem[\protect\citeauthoryear{Bressan et al.}{2012}]{bre12}
	Bressan A., Marigo, P., Girardi L., Salasnich, B., Dal Cero, C., Rubele, S., Nanni, A., 2012, \mnras, 427, 127	
	
	\bibitem[Brown et al.\ (2018)]{bro18} Brown, A. G. A., Vallenari, A., Prusti, T. et al., 2018, \aap, 616, 1G 
	
	\bibitem[Brown et al.\ (2020)]{bro20} Brown, A. G. A., Vallenari, A., Prusti, T. et al., 2020, 2020arXiv201201533G 
	
		
	
	\bibitem[Brunthaler et al.\ (2011)]{bru11} Brunthaler, A., Reid, M.J., Menten, K.M., et.al. 2011, AN, 332, No.5, 461
	
	
	\bibitem[Bukowiecki et al.\ (2011)]{buk11} Bukowiecki, L., Maciejewski, G., Konorski, P., Strobel, A., 2011, AcA 61, 231
	
	\bibitem[Cantat-Gaudin et al.\ (2018)]{can18} Cantat-Gaudin, T., Jordi, C., Vallenari, A., Bragaglia, A.,  Balaguer-Nunez, L., Soubiran, C., Bossini, D., Moitinho, A., Castro-Ginard, A., Krone-Martins, A., and 3 coauthors., 2018, \aap, 618, 93
	
	\bibitem[Cantat-Gaudin et al.\ (2020)]{can20} Cantat-Gaudin, T., Anders, F., Castro-Ginard, A., Jordi, C., Romero-Gómez, M., Soubiran, C., Casamiquela, L., Tarricq, Y., Moitinho, A., Vallenari, A.,  Bragaglia, A., Krone-Martins, A., Kounkel, M.,	2020, \aap, 640, 1
	
	\bibitem[Carney (2001)]{car01} Carney, B. 2001, Star Clusters, Saas-Fee Advanced Course 28, Lecture Notes 1998, Swiss Society for Astrophysics and Astronomy, eds.\ L.~Labhardt and B.~Binggeli
	(Berlin:  Springer-Verlag) pp.~1-222
	
	\bibitem[Carraro et al.\ (2010)]{car10} Carraro, G., Costa, E., Ahumada, J.A., 2010, \aj, 140, 954 
	
	\bibitem[Chen and Han (2009)]{che09} Chen, X. F. and Han, Z. W., 2009, \mnras, 395, 1822
	
	\bibitem[Chen  et al.\ (2017)]{che17} Chen, X., de Grijs, R., Deng, L., 2010, \mnras, 464, 119 
		
	\bibitem[\protect\citeauthoryear{Chiosi et al.}{1992}]{chi92}
	Chiosi, C., Bertelli, G., Bressan, A., 1992, \araa, 30, 235
	
	\bibitem[Dutra et al.(2002)]{dut02} Dutra, C.M., Santiago, B.X., Bica, E., 2002, \aap, 383, 219
	
	
	\bibitem[\protect\citeauthoryear{Ekstr\"om et al.}{2012}]{eks12}
	Ekstr\"om, S., Georgy, C., Eggenberger, P., Meynet, G., Mowlavi, N., Wyttenbach, A., Granada, A., Decressin, T., Hirschi, R., Frischknecht, U., Charbonnel, C., Maeder, A., \aap, 2012, 537,146 (E12)
	
	\bibitem[Fernie \  (1963)]{fer63} Fernie, J.D., 1963, \aj, 68, 780
	
	
	
	\bibitem[Ferraro \ (2016)]{fer16} Ferraro, F.R., 2016, Star Clusters and Black Holes in Galaxies across Cosmic Time, Proceedings of the International Astronomical Union, IAU Symposium, Volume 312, pp. 171-180.
	
	
	\bibitem[Fiorentino et al.\  (2014)]{fio14} Fiorentino, G., Lanzoni, B., Dalessandro, E., et al.,2014, \apj, 783, 34
	
  \bibitem[Fitzgerald \ (1970)]{fit70} Fitzgerald, M.P., 1970, \aap, 4, 234
	
	
	\bibitem[Gilliland et al.\  (1998)]{gil98} Gilliland, R. L., Bono,G., Edmonds, P. D., et al.,1998, \apj, 507, 818
	
	\bibitem[Girardi et al.\ (2002)]{gir02} Girardi, L., Bertelli, G., Bressan, A., Chiosi, C., Groenewegen, M.A.T., Marigo, P., Salasnich, B., Weiss, A., 2002, \aap, 391, 195
	
	\bibitem[Georgy et al.\ (2013)]{geo13} Georgy, C.,Ekstr\"om, S., Granada, A., Meynet, G., Mowlavi, N., Eggenberger, P., Maeder, 2013, \aap, 553, 24 (G13)
	
	
	\bibitem[\protect\citeauthoryear{Guetter \& Vrba}{1989}]{gue89}
	Guetter, H. H., \& Vrba, F. J. 1989, \aj, 98, 611
	
	
	\bibitem[Hills and Day (1976)]{hil76} Hills, J. and Day, C.,1976, Astron. Lett., 17, 87
	
	
	\bibitem[Johnson and Soderblom (1987)]{joh87} Johnson, D. R. H., \& Soderblom, D. R., 1987, \aj, 93, 864
	
	\bibitem[Kepley et al.\  (2007)]{kep07} Kepley, A., Morrison, H.L., Helmi, A., Kinman, T.D., et al., 2007, \aj, 134, 1579
	
	
	\bibitem[\protect\citeauthoryear{Kharchenko et al.}{2013}]{kha13}
	Kharchenko, N.V., Piskunov, A.E., Schilbach, E., R{\"o}ser, S., Scholz,
	R.D., 2013, \aap, 558, 53
	
	\bibitem[\protect\citeauthoryear{Kilkenny et al.}{1998}]{kil98}
	Kilkenny, D., Van Wyk, F., Roberts, G., Marang, F., Cooper, D., 1998, \mnras, 294, 93
	
	\bibitem[King \ (1966)]{kin66} King, I., 1966, \aj, 71, 64
	
	\bibitem[Koornneef \ (1983)]{koo83} Koornneef, J., 1983, \aap, 128, 84
	
	
	
	\bibitem[Lazovik and Rastorguev \ (2020)]{laz20} Lazovik, Y.A., Rastorguev, A.S., 2020, \aj, 160, 2020
	
	\bibitem[Li et al.\ (2017)]{li17} Li, C., de Grijs R., Deng L., Milone A. P., 2017, \apj, 844, 119
	
	\bibitem[\protect\citeauthoryear{Lim et al.}{2009}]{lim09}
	Lim, B., Sung, H., Bessell, M. S., Karimov, R.,Ibrahimov, M. 2009, JKAS, 41, 161
	
	\bibitem[Lim et al.\ (2014)]{lim14} Lim, B., Sung, H., Kim, J.S.,  Bessell, M., Park, B-G., 2014, \mnras, 443, 454
	
	\bibitem[\protect\citeauthoryear{Lim et al.}{2016}]{lim16}
	Lim, B., Sung, H., Kim, J.S., Bessell, M.S., Hwang, N., Park, B-G. 2016, \apj, 831, 116
	
	\bibitem[Lim et al.\ (2019)]{lim19} Lim, B., Rauw, G., Nazé, Y., Sung, H., Hwang, N., Park, B-G.,  2019, NatAs, 3, 76	
	
	
	\bibitem[Lindegren et al.\ (2018)]{lin18} Lindegren, L., Hernandez, J., Bombrun, A., Klioner, S. et al., 2018, \aap, 616, A2 
	
	\bibitem[Lindegren et al.\ (2020)]{lin20} Lindegren, L., Hernandez, J., Bombrun, A., Klioner, S. et al., 2020, 2020arXiv201201533G  
	
	\bibitem[\protect\citeauthoryear{Lohr et al.}{2018}]{loh18}
	Lohr, M. E., Negueruela, I.,  Tabernero, H. M.,  Clark, J. S.,  Lewis, F. and  Roche, P. 2018, \mnras, 478, 3825 (L18)  
	
	
	\bibitem[Lombardi et al.\  (2002)]{lom02} Lombardi, Jr., J. C., Warren, J. S., Rasio, F. A., Sills, A., and Warren, A. R., 2002, \apj, 568, 939
	
	\bibitem[\protect\citeauthoryear{Maciejewski \& Niedzielski}{2007}]{mac07} 
	Maciejewski, G., \& Niedzielski, A. 2007,\aap, 467, 1065
	
	\bibitem[McCrea \ (1964)]{mc64} McCrea, W. H., 1964, \mnras, 128, 147
	
	\bibitem[\protect\citeauthoryear{Marco et al.}{2014}]{mar14}
	Marco, A., Negueruela, I., Gonzalez-Fernandez, C., Maiz Apellaniz, J., Dorda, R., Clark, J.S., 2014, \aap, 567, 73
	
	\bibitem[\protect\citeauthoryear{Marigo et al.}{2008}]{mar08}
	Marigo, P., Girardi, L., Bressan, A., et al., 2008, \aap, 482, 883
	
	\bibitem[\protect\citeauthoryear{Mermilliod}{1981}]{mer81}
	Mermilliod, J.-C., 1981, \aap, 97, 235
	
	\bibitem[\protect\citeauthoryear{Menzies et al.}{1991}]{men91}
	Menzies, J. W., Marang, F., Laing, J. D., Coulson, I. M., Engelbrecht, C. A., 1991, \mnras, 248, 642
	
	\bibitem[\protect\citeauthoryear{Molina Lera et al.}{2018}]{mol18}
	Molina Lera, J.A, Baume, G., Gamen, R., 2018, \mnras, 480, 2386
	
	
	\bibitem[\protect\citeauthoryear{Negueruela \& Marco}{2012}]{neg12}
	Negueruela, I., \& Marco, A. 2012, \aj, 143, 46 (N12)
	
	\bibitem[\protect\citeauthoryear{Negueruela et al.}{2016}]{neg16}
	Negueruela, I., Clark, J. S., Dorda, R., González-Fernández, C., Marco, A., Monguio, M., 2016,
	Vol.507, page 75 , Edited by Ian Skillen, Marc Balcells, and Scott Trager. ASP Conference Series, San Francisco: Astronomical Society of the Pacific (ASPC).
	
	\bibitem[\protect\citeauthoryear{Negueruela et al.}{2018}]{neg18}
	Negueruela, I., Monguio, M., Marco, A., Tabernero, H.M., Gonzalez-Fernandez, C.,  Dorda, R.,  2018, \mnras, 477,2976
	
	
	\bibitem[Sampedro et al.\ (2017)]{sam17} Sampedro, L., Dias, W.S., Alfaro, E.J., Monteiro, H., Molina, A., 2017, \mnras, 470, 3937 
	
	\bibitem[Sandage \ (1953)]{san53} Sandage, A.R., 1953, \aj, 58, 61
	
	\bibitem[Schmidt-Kaler\ (1982)]{sch82} Schmidt-Kaler, Th.\ 1982, in Landolt-Bornstein, Numerical Data and Functional Relationships in Science and Technology, New Series, Group VI, Vol.2b, eds.\ K.~Schaifers \& H.~H. Voigt (Berlin:  Springer), p.~14 (SK82)
	
	
	\bibitem[Sch{\"o}nrich and Binney (2010)]{sch10} Sch{\"o}nrich R., Binney J., Dehnen, W.,  2010, \mnras, 403, 1829
	
	
	\bibitem[\protect\citeauthoryear{Sung et al.}{1997}]{sun97}
	Sung, H.,  Bessell, M.S., See-Woo, L., 1997, \aj, 114, 2644
	
	Sung, H., \& Bessell, M. S. 1999, \mnras, 306, 361
	
	
	
	\bibitem[\protect\citeauthoryear{Sung et al.}{2013}]{sun13}
	Sung, H., Lim, B., Bessell, M.S. Kim, J.S., Hur, H., Chun, M-Y., Park,
	B-G., 2013, JKAS, 46, 97 (S13)
	
	
	
	\bibitem[\protect\citeauthoryear{Tadross}{2008}]{tad08}
	Tadross, A. L. 2008, \mnras, 389, 285
	
	\bibitem[Tout et al.\  (1997)]{tou97} Tout, C. A., Aarseth, S. J., Pols, O. R., Eggleton, P. P., 1997, \mnras, 291, 732
	
	

\end{theunbibliography}
\end{document}